\documentclass[reprint,superscriptaddress,amsmath,amssymb,prl,floatfix]{revtex4-2}
\usepackage[dvipdfmx]{graphicx}
\usepackage{dcolumn}
\usepackage{bm}
\usepackage{xcolor}
\usepackage{here}

\usepackage{amsthm}
\theoremstyle{definition}

\begin{document}
\title{Designing robust trajectories by  lobe dynamics \\ in low-dimensional Hamiltonian systems}

\author{Naoki Hiraiwa}
\affiliation{Department of Aeronautics and Astronautics, Kyushu University, 744 Motooka, Nishi-ku, Fukuoka 8190395, Japan}
\affiliation{Colorado Center for Astrodynamics Research, University of Colorado Boulder, 3775 Discovery
Drive, Boulder, Colorado 80303, USA}
\author{Mai Bando}
\affiliation{Department of Aeronautics and Astronautics, Kyushu University, 744 Motooka, Nishi-ku, Fukuoka 8190395, Japan}
\author{Isaia Nisoli}
\affiliation{Instituto de Matem{\'a}tica, Universidade Federal do Rio de Janeiro, Avenida Athos da Silveira Ramos, 149, Edif{\'i}cio do Centro de Tecnologia, Bloco C (T{\'e}rreo), Cidade Universit{\'a}ria 21941-909, Brazil}
\author{Yuzuru Sato}
\email{ysato@math.sci.hokudai.ac.jp}
\affiliation{RIES-MSC/Department of Mathematics, Hokkaido University, N12 W7 Kita-ku, Sapporo, 0600812 Hokkaido, Japan}
\affiliation{London Mathematical Laboratory, 14 Buckingham Street, London WC2N 6DF, United Kingdom}
\date{\today}

\begin{abstract}
Modern space missions with uncrewed spacecraft require robust trajectory design to connect multiple chaotic orbits by small controls. 
To address this issue, we propose a control scheme to design robust trajectories by leveraging a geometrical structure in chaotic zones, known as a {\it lobe}. Our scheme shows that appropriately selected lobes reveal possible paths to traverse chaotic zones in a short time. The effectiveness of our method is demonstrated through trajectory design in both the standard map and Hill’s equation.
\end{abstract}

\maketitle

The Artemis program~\cite{creech2022artemis}, including the uncrewed cargo mission to the lunar Gateway~\cite{ikenaga2020study}, has attracted significant attention from aerospace engineers. This mission demands frequent transportation from the Earth to the lunar Gateway, emphasizing the need for a method to design a robust transfer. Additionally, many recent deep-space missions aimed at enhancing our knowledge of planetary science~\cite{Bandyopadhyay2016review} have utilized small satellites with limited fuel and maneuver capabilities. In such modern space missions with uncrewed spacecraft, trajectory design must incorporate chaotic orbits because spacecraft are expected to traverse chaotic zones to reach the Moon under severe thrust and transfer time conditions. 
To address these issues, we propose a control scheme to design robust trajectories by leveraging a geometrical structure in chaotic zones, known as a {\it lobe}~\cite{rom1990transport}.

Conventionally, spacecraft trajectories affected by the gravity of celestial bodies have been designed by connecting paths near tori in Hamiltonian systems with adequate controls. In the two-body problem, optimal trajectories are formed based on Hohmann transfer---the minimum-fuel two-impulsive transfer between coplanar circular orbits, or flyby, which is a gravity-assist maneuver by a planet~\cite{prussing1993orbital}. 
For the restricted three-body problems, several effective techniques have been studied, including \textit{tube dynamics}, which constitutes the transport structure of cylindrical invariant manifolds~\cite{koon2000heteroclinic,rossDynamicalSystems}, ballistic lunar transfers, which efficiently utilize solar forces~\cite{parker2014,MCCARTHY2023556}, and resonant gravity assist, which consists of multiple flybys around the same planet~\cite{ross2003design}. 

The primary obstacle in trajectory design is to handle chaotic orbits.
In the literature on nonlinear dynamics, the concept of controlling chaos~\cite{ott1990controlling, pyragas1992continuous} focuses on stabilizing chaotic motion through small perturbations to the system. Conversely, harnessing chaos~\cite{yamaguti1994towards,jaeger2004harnessing} attempts to exploit the characteristics of chaotic motion, including so-called 
\textit{targeting}~\cite{shinbrot1990using}, 
where the sensitivity to initial conditions is used to swiftly direct the system to a desired point in state space.
The trajectory design in the Earth--Moon system has been studied following the seminal work by Bollt and Meiss~\cite{bollt1995targeting}, which introduced an approach to shorten a natural chaotic transfer trajectory by leveraging recurrence and instability. Subsequently, Schroer and Ott developed the pass targeting method~\cite{schroer1997targeting}. 
Another research direction involves trajectory design based on Lagrangian coherent structures. This line of research focuses on adding small controls to get over a separatrix between different coherent structures in fluid dynamics~\cite{senatore2008fuel,ramos2018lagrangian,krishna2022finite,haller2023transport}. These methods are similar to the control techniques based on tube dynamics in spacecraft trajectory design~\cite{koon2000heteroclinic}.

In this Letter, we present a control method to design robust trajectories based on lobe dynamics, which is a finer geometrical structure than tube dynamics. Although lobe dynamics has been studied to analyze transport in dynamical systems~\cite{duan1997lagrangian,coulliette2001intergyre,Lekien2007lagrangian,oshima2015jumping,naik2017computational}, it has not been used for trajectory design. 
 We establish a framework to design robust trajectories to connect start and goal orbits via a few chaotic orbits within  selected lobes. As a result, 
we notably find  shorter-time transfers than those in the previous work in the Earth--Moon system~\cite{bollt1995targeting,schroer1997targeting}.

To design finite-time trajectories in Hamiltonian systems, we presume knowledge regarding the equation of motion and the instant measurements of the spacecraft's position and velocity. Moreover, we assume that the trajectories remain in the same energy surface in a Hamiltonian system before and after control.   
Our investigation is focused on a specific finite-time trajectory, departing from a start orbit in an elliptic island and arriving at a goal orbit in another elliptic island in a low-dimensional Hamiltonian system.

We illustrate our scheme in the standard map and Hill’s equation. The former demonstrates a simple example for our method, while the latter applies it to a more realistic scenario. The standard map~\cite{CHIRIKOV1979263} serves as the simplest model of Hamiltonian systems, expressed as  
\begin{align}
    p_{n+1}&=p_n+K\sin \theta_n, \nonumber\\
    \theta_{n+1}&=\theta_n+p_{n+1}
    ~~(\mbox{mod} ~2\pi), \label{eq:standard_map}
\end{align}
where the Hamiltonian of the flow is given as $H(p,\theta,t) = p^2/2 + K\cos\theta\sum_{n=-\infty}^\infty\delta(t-n)$. We set $K = 1.2$ as an example, which gives non-integrable chaotic dynamics for many orbits with a positive top Lyapunov exponent. 
Hill's equation~\cite{szebehely1967theories} is a non-dimensional model for the Earth--Moon planar circular restricted three-body problem, expressed as
\begin{align}
    \ddot{x} - 2\dot{y} - x &= - \frac{(1-\mu)(x+\mu)}{{r_1}^3} - \frac{\mu(x-1+\mu)}{{r_2}^3}, \nonumber\\
    \ddot{y} + 2\dot{x} - y &= - \frac{(1-\mu)y}{{r_1}^3} - \frac{\mu y}{{r_2}^3},
    \label{eq:EoM_spacectaft}
\end{align}
where the position of a spacecraft is $(x,\,y)$, $r_1 = \sqrt{(x+\mu)^2+y^2}$, $r_2 = \sqrt{(x-1+\mu)^2+y^2}$, and the masses for the Earth and Moon are given as $1-\mu$ and $\mu$, respectively. In these standard coordinates, the unit of length is the distance between the Earth and Moon given as $3.844\times 10^5$~(km), the unit of mass is the sum of the masses of the Earth and Moon as $6.046\times 10^{24}$~(kg), and the unit of time is the inverse of the rotation rate in the system as  $1.042$~(h).
The Jacobi integral given by 
\begin{equation*}
    J = x^2 + y^2 + 2\frac{1-\mu}{r_1} + 2\frac{\mu}{r_2} + \mu(1-\mu) - (\dot{x}^2 + \dot{y}^2)
\end{equation*}
restricts the flow to a three-dimensional subspace in the four-dimensional state space. We set $\mu = 1.21509\times10^{-2}$ and $J=3.16$, which possesses sufficient energy level to enable transfers from the Earth to the Moon and indicates chaotic dynamics.

\begin{figure}[t]
    \centering
    \includegraphics[width=0.9\columnwidth]{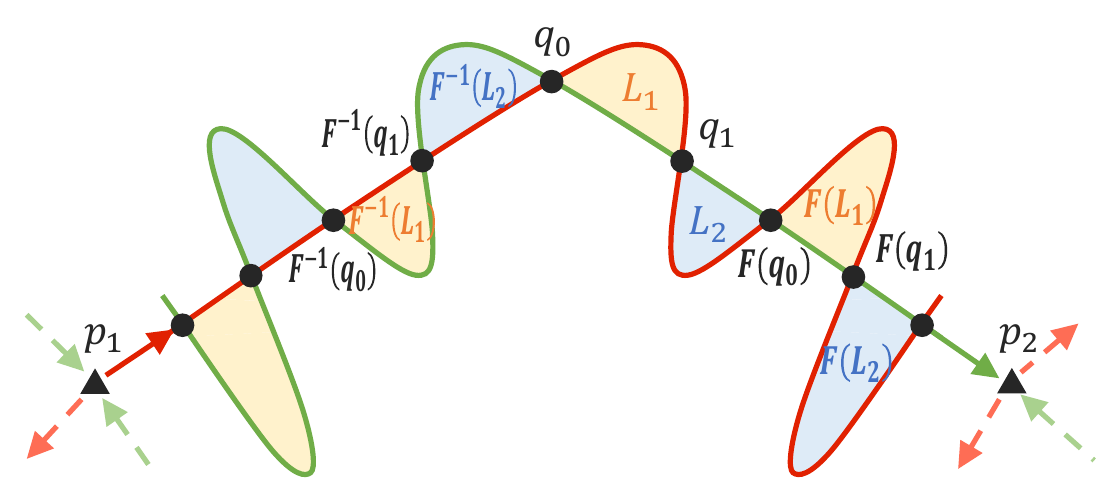}
    \caption{\label{fig:lobe} Lobes $L_1, L_2$, and their transport by an area- and orientation-preserving map $F$. Hyperbolic periodic points, $p_1$ and $p_2$, and intersection points are denoted as triangles and black dots, respectively. The stable and unstable manifolds associated with the periodic points are depicted by green and red lines, respectively. Yellow and blue regions represent two different lobe sequences.}%
\end{figure}

The key concept in our control scheme is lobe dynamics~\cite{rom1990transport}, which was initially conceptualized for analyzing phase space volume transportation in Hamiltonian systems~\cite{rom1990analytical}. The lobes are identified by finding two hyperbolic periodic points, $p_1$ and $p_2$, within the two-dimensional state space of an area- and orientation-preserving map $F$. In chaotic systems, the stable and unstable manifolds of $p_1$ and $p_2$ can intersect infinitely many times if they lie in the same chaotic zone.
By identifying two adjacent intersection points, $q_0$ and $q_1$, the region enclosed by segments of the stable and unstable manifolds between these points constitutes a lobe~\cite{rom1990transport}. A \textit{lobe sequence} is defined as a series of lobes mapped by $F$. Each pair of stable/unstable manifolds may form multiple lobe sequences. Figure~\ref{fig:lobe} schematically illustrates two such sequences originating from $L_1$ and $L_2$, mapped by $F$. 
Because one lobe is mapped to another by $F$, the trajectories starting within the same lobe exhibit similar behavior over a finite time and are encircled by invariant manifolds associated with unstable periodic points. This characteristic makes lobes suitable for robust trajectory design in our scheme. However, lobes, being infinite in number, eventually fold intricately to become dense in the chaotic zone. To leverage lobe sequences for robust trajectory design, we provide the definition of an \textit{effective lobe sequence}. 
Let $B_{\varepsilon}(c)$ be $\varepsilon$-ball with the center $c$ in a lobe. 
The radius $r_L$ of a lobe $L$ is defined as the largest $\varepsilon$ in all possible $B_{\varepsilon}(c)$'s in the lobe; $r_{L}:= \max_{ c\in L, B_{\varepsilon}(c)\subset L} \varepsilon$. 
As the mapping iterates, lobes are typically stretching out with a positive Lyapunov exponent, and the sequence of the radii of the lobes asymptotically converges to $0$. Therefore, an effective lobe sequence is defined as a lobe sequence composed of a finite number of the lobes whose radius is larger than a minimum lobe radius $r^*$. For practical computations, the radius $r_{L}$ is estimated as the Hausdorff semi-distance between the lobe's center of gravity and its boundary. This radius indicates allowable observational/operational error bounds during a transfer. 
\begin{figure}[t]
    \centering
    \includegraphics[width=\columnwidth]{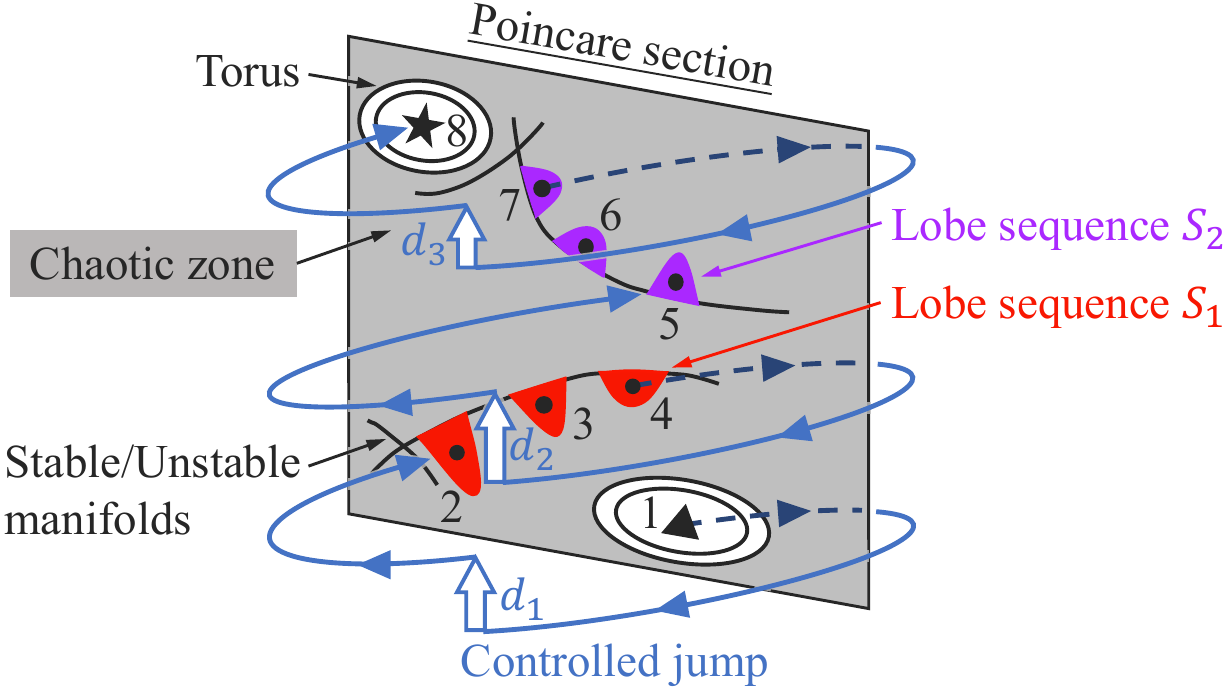}
    \caption{\label{fig:control_scheme} Schematic view of our control. The start point, goal point, and centers of gravity of the lobes are denoted as a triangle, a star, and black dots, respectively.  Finitely long effective lobe sequences $S_i\ (i=1,2)$ are used for the transfer. In this example, three controlled jumps outside the Poincar\'{e} section are required to connect two orbits in different elliptic islands. The numbers indicate the order of the transfer. The total control cost $D=\sum_{k=1}^N d_k$ ($N=3$ here) is minimized.}
\end{figure}

Figure~\ref{fig:control_scheme} outlines our control scheme, which establishes a start point on the start orbit $O_s$ and a goal point on the goal orbit $O_g$ to construct the desired orbit-to-orbit transfer with the smallest total control cost. The definition of control costs depends on application. Given a trajectory connecting $O_s$, effective lobe sequences $S_i$ ($i = 1,2,\ldots,N-1$), and $O_g$ by external controls, this trajectory becomes a robust transfer bounded by the segments of stable and unstable manifolds.
In our scheme, this designed trajectory first jumps from a start point on $O_s$ to a point on $S_1$ at a control cost $d_1$. The trajectory remains within $S_1$ without control before the next jump to $S_2$ at a cost $d_2$. Controlled jumps from $S_i$ to $S_{i+1}$ are repeated until the trajectory reaches $O_g$ by the $N$-th jump at a cost $d_N$. These controlled jumps between lobe sequences help overcome the partial barriers formed by the boundaries of resonances~\cite{MACKAY19871,meiss2015thirty} or cantori~\cite{MACKAY1984transport}. Although the trajectory may remain within the same lobe sequence for an extended period without control, it is necessary to jump to another effective lobe sequence within a finite time for maintaining system controllability, because the radius of a lobe eventually converges to zero. Thus, a small number of the selected lobe sequences contribute to short-time transfers.  The control costs $\{d_k\}$ are determined under the following constraints:
\begin{enumerate}
    \item The trajectory moves to the center of gravity of an effective lobe with $r_L > r^*$ or a goal point on $O_g$ by control.
    \item The cost of each jump $d_k$ satisfies $0 < d_k < d^*$. 
    \item The trajectory remains within an effective lobe sequence for at least one step.
    \item Minimize the total cost $D = \sum_{k=1}^N d_k$.
\end{enumerate}
The maximum control cost $d^*$ represents the maximum thrust of the engines at one step.
Before optimization, we predetermine potential start points on $O_s$, potential goal points on $O_g$, and the constraint parameters $r^*$ and $d^*$. To select effective lobe sequences, we first select candidates for the first effective lobe sequence $S_1$ that can be reached from the potential start points on $O_s$ by a controlled jump with $d_1<d^*$. Similarly, we then explore candidates for $S_2$ reachable from $S_1$.  This procedure concludes with finding candidates for the final effective lobe sequence accessible to the goal points on $O_g$ by a controlled jump with $d_N<d^*$. 
For any pair of $O_s$ and $O_g$ in different elliptic islands, if there exists a chaotic zone between them, we can find lobe sequences by using a sufficiently small $r^*$ and sufficiently large $d^*$ given that a lobe sequence can approach any elliptic islands in the long run. 
The optimization is performed for finite combinations of jumps among the start points, a few effective lobe sequences, and the goal points. A larger $r^*$ and smaller $d^*$ contribute to reducing the computational cost for the optimization. A detailed explanation of this optimization is presented in the Supplemental Material~\cite{SM}.

\begin{figure}[t]
    \centering
    \includegraphics[width=0.9\columnwidth]{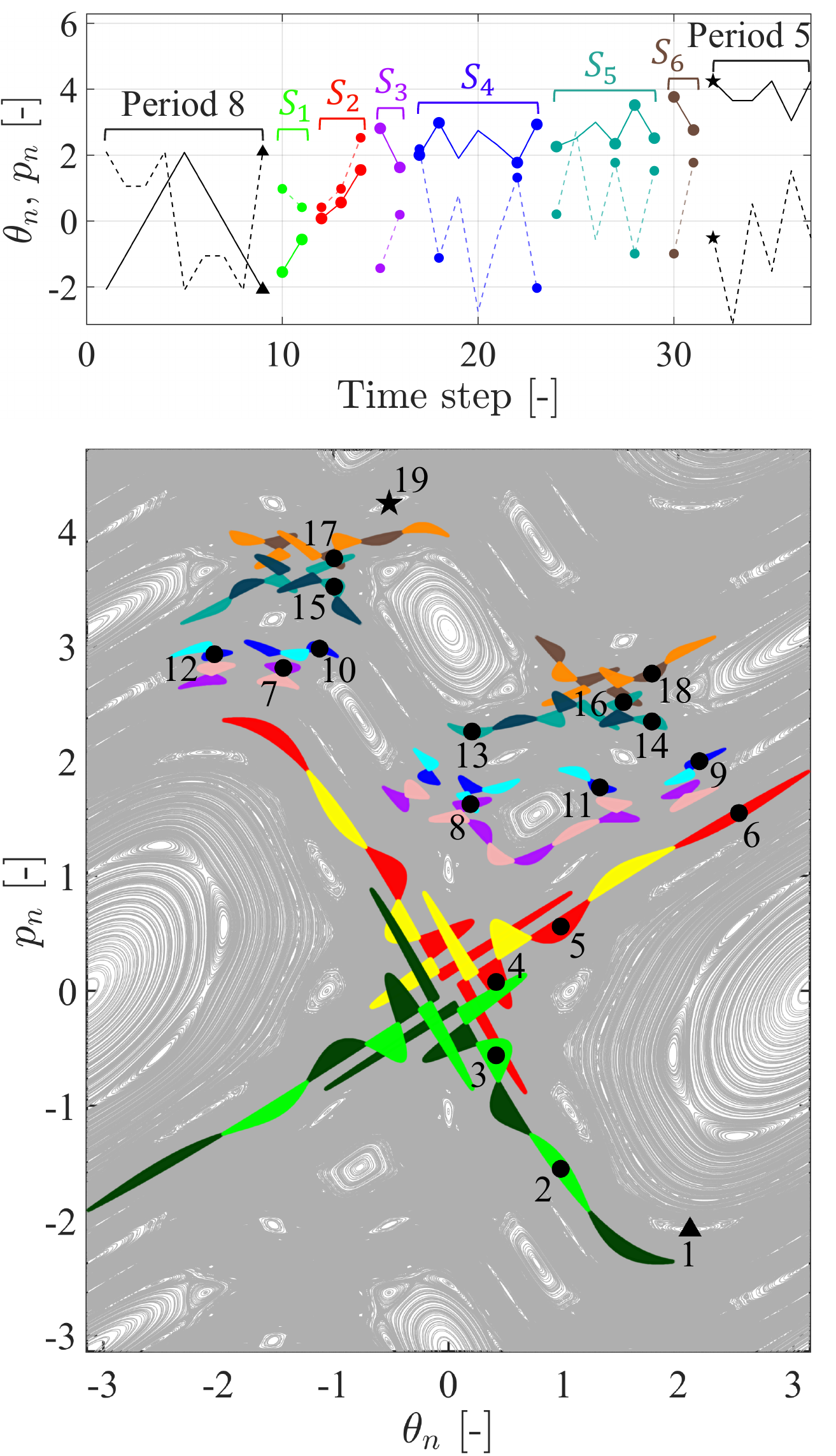}
    \caption{Optimal trajectory for the standard map with $K=1.2$, with a total cost $D=2.1333$ and transfer time $\Delta n = 23$, where $r^*=0.02$ and $d^*=0.64$: Controlled time series of $p_n$ (solid line) and $\theta_n$ (dashed line) (top), and state space of the standard map (bottom) are depicted. Gaps on the top panel indicate controlled jumps. Different effective lobe sequences are colored differently. The start point, goal point, and centers of gravity of the adopted lobes are denoted as a triangle, a star, and dots, respectively. The numbers in the bottom panel represent the order of transfer, similar to those in Fig.~\ref{fig:control_scheme}.}
    \label{fig:time_history}
\end{figure}

We first apply our control scheme to the standard map, given by the stroboscopic map of the kicked rotator. Starting with $(p_n,\,\theta_n)$ on the Poincar{\'e} section at time $t = n$, the momentum changes to $p_{n}+K\sin\theta_n$,  and then is adjusted to $p_{n}+K\sin\theta_n+\Delta p_n$ by a control at time $t=n+\eta_n\,(0<\eta_n<1)$. 
The position $\theta_n$ is integrated with the modified momentum after $t=n+\eta_n$. 
Upon returning to the Poincar{\'e} section at time $t=n+1$, the controlled jump to $(p'_{n+1},\,\theta'_{n+1})$ is established as 
\begin{align}
    p'_{n+1} &= p_n + K\sin\theta_n + \Delta p_n = p_{n+1} + \Delta p_n, \nonumber\\
    \theta'_{n+1} &= \theta_n + p_{n+1} + (1-\eta_n)\Delta p_n \nonumber\\
                  &= \theta_{n+1} + (1-\eta_n)\Delta p_n.
    \label{eq:jump}
\end{align}
Because the combinations of $(p_n,\,\theta_n)$ and $(p'_{n+1},\,\theta'_{n+1})$ are given in the optimization process, we can compute the control parameters $\Delta p_n$ and $\eta_n$ from Eq.~\eqref{eq:jump}.
Within this control framework, each control cost is quantified as $d_n=|\Delta p_n|$, subject to $|\theta_{n+1}'-\theta_{n+1}|<|p_{n+1}'-p_{n+1}|<d^*$.
In this example, the start and goal orbits, $O_s$ and $O_g $, are selected as periodic orbits with periods $8$ and $5$, respectively.
All points on each periodic orbit are regarded as potential start/goal points. The minimum lobe radius and maximum jump cost are given as $r^*=0.02$ and $d^*=0.64$, respectively. 
To designate intermediate waypoints, we select $12$ effective lobe sequences with up to $9$ step length.
Thus, our scheme finds the optimal trajectory with a total cost $D=2.1333$ and transfer time $\Delta n = 23$, including the coasting time within lobe sequences without controls, as illustrated in Fig.~\ref{fig:time_history}. As a result, our optimal trajectory achieves a short-time transfer. The total cost is larger than the minimum cost of a direct jump from a start point to a goal point ($d_1=0.9673$), due to the necessity of multiple jumps under the constraint $d_k < d^* = 0.64$.
The colored regions in the lower panel represent the selected $12$ effective lobe sequences, from which six effective lobe sequences $S_1,\ldots,S_6$ are adopted, corresponding to colored lines in the upper panel. The gaps in the upper panel signify the controlled jumps.

\begin{figure}[t]
    \centering
    \includegraphics[width=0.95\columnwidth]{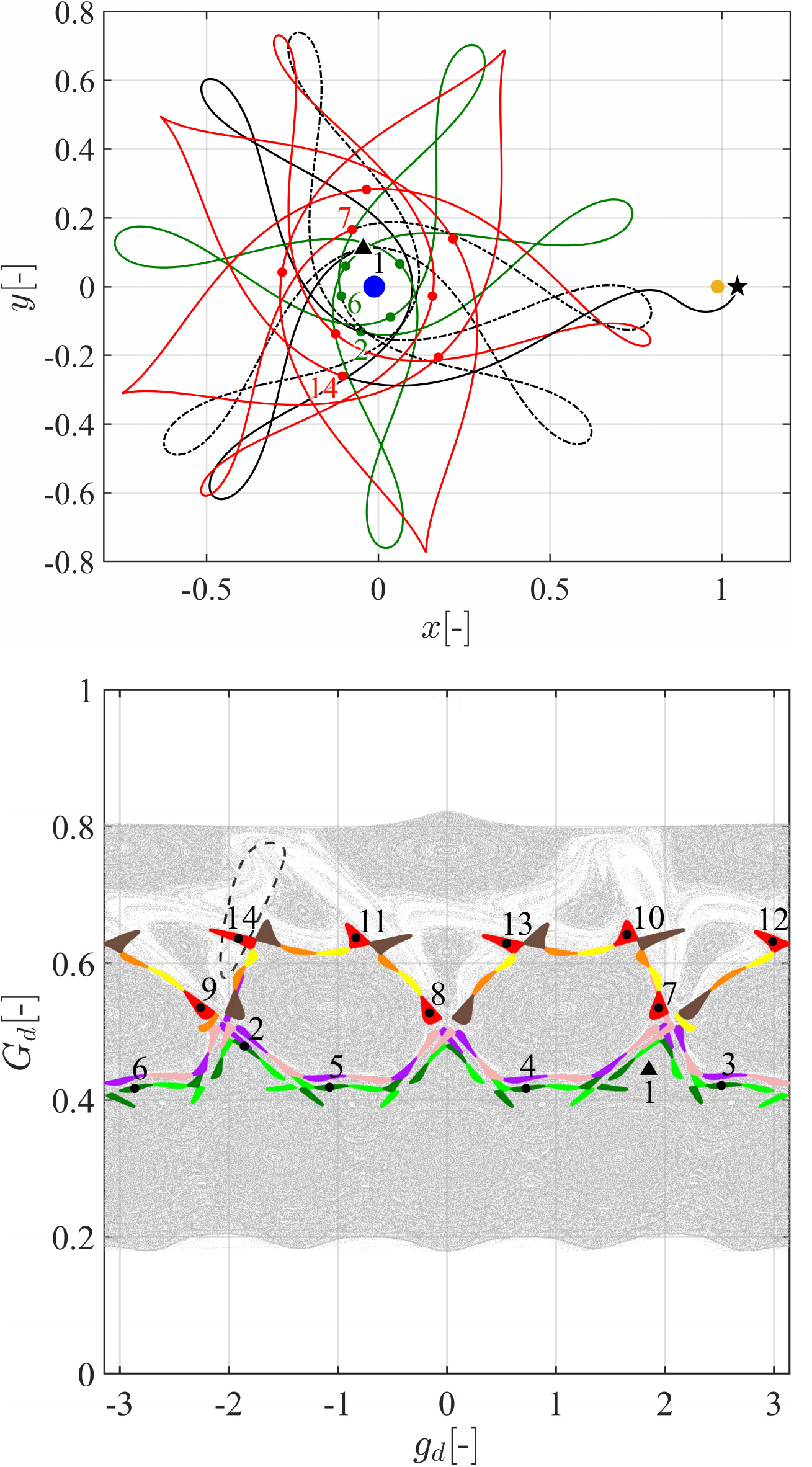}
    \caption{Optimal trajectory for the Hill's equation with $J=3.16$, with a total cost $D = 0.1511$ ($154.7849$~[m/s]) and transfer time $\Delta t=37.5827$ [$163.2018$ (days)], where $r^*=0.002$ and $d^* = 0.09760$ [$100$~(m/s)]: Controlled trajectory in the position space (top) and Poincar{\'e} section at perigee passage  (bottom) are depicted. A dash-dotted line denotes the controlled transition between effective lobe sequences, and solid lines represent the other part of the transfer. Blue and yellow dots indicate the Earth and Moon, respectively. The region surrounded by a dashed line in the Poincar{\'e} section is the gate to the Moon realm. 
    Other notations are the same as those in Fig.~\ref{fig:time_history}.}
    \label{fig:transfer_CR3BP}
\end{figure}
Similarly, we implement our control scheme for Hill's equation with $J=3.16$ by utilizing the Poincar{\'e} section at perigee passage. The control cost $d_k$ is set as the magnitude of impulsive velocity change at a control point outside the Poincar{\'e} section. The controlled jumps in this optimization only change the velocity direction so that the Jacobi integral remains the same.
We facilitate a robust transfer originating from one of the periapses of the $7$:$2$ neutrally stable resonant orbit and leading to the section at $y = 0$ and $\dot{y} > 0$ within the Moon realm, by focusing on $8$ effective lobe sequences. The constraint parameters are set as $r^*=0.002$ and  $d^* = 0.09760$ [$100$~(m/s)]. 
The derived optimal trajectory is illustrated in Fig.~\ref{fig:transfer_CR3BP},  and is characterized by a total cost $D=0.1511$ [$154.7849$~(m/s)] and transfer time $\Delta t=37.5827$ [$163.2018$~(days)], including the coasting time on lobe sequences without control and continuous trajectory outside the Poincar{\'e} section until $y = 0\,(\dot{y} > 0)$ around the Moon.
The lower panel of Fig.~\ref{fig:transfer_CR3BP} shows the Poincar{\'e} section at perigee passage. This Poincar{\'e} section is rendered in action $G_d$ - angle $g_d$ coordinates by translating the spacecraft's state at a perigee into the canonical variables known as Delaunay elements.
An increase in $G_d$ at perigees typically indicates a larger distance between the Earth and spacecraft compared to previous positions. A detailed explanation is given in the Supplemental Material~\cite{SM}. The region enclosed by the dashed line in the lower panel of Fig.~\ref{fig:transfer_CR3BP} signifies the stable manifold of the Lyapunov orbit, acting as the sole control-free path from the Earth realm to the Moon realm, which corresponds to the goal orbit $O_g$ in our scheme. The colored regions in the lower panel represent the eight selected effective lobe sequences, from which two effective lobe sequences $S_1$ and $S_2$ are adopted, corresponding to colored lines in the upper panel. The transfer time of our optimal trajectory is much shorter than that of the Bollt and Meiss trajectory [$748$~(days)]~\cite{bollt1995targeting} and that of the Schroer and Ott trajectory [$293$~(days)]~\cite{schroer1997targeting}, despite our trajectory starting farther away from the Moon.
According to Ref.~\cite{ikenaga2020study}, the cargo transport to the Moon may have a total cost of $0$--$400$~(m/s) and transfer time of several days to $1$ year, suggesting that our result of $D=154.7849$~(m/s) and $\Delta t=163.2018$~(days) is practically useful for the preliminary trajectory design in the Earth--Moon system.

In summary, we propose a control scheme to design robust trajectories utilizing effective lobe sequences, making the trajectories insensitive to external perturbations. Our scheme reveals that the effective lobes can indicate possible paths to traverse chaotic zones in a short time, with small controls, and with limited fuel. 
The examples with the standard map and Hill's equation demonstrate that our control scheme can construct trajectories with a short transfer time by leveraging lobe dynamics. 
In Hamiltonian systems with three or more degrees of freedom, tori may not impede dynamics, which allows the trajectories to migrate from the inside of a torus to the outside. On the other hand, the literature of Refs.~\cite{WIGGINS1990471,dellnitz2005transport} suggests that stable/unstable manifolds associated with normally hyperbolic invariant manifolds may form lobes in high-dimensional Hamiltonian systems. Thus, the control based on lobes in high-dimensional Hamiltonian systems remains as a challenging future work.

\begin{acknowledgements}
    The authors thank Prof. J. D. Meiss and Prof. D. J. Scheeres for their insightful discussions. N.H. acknowledges the support of Japan Society for the Promotion of Science (JSPS) KAKENHI Grant No. JP 23KJ1692. 
    M.B. was supported by Japan Science and Technology Agency (JST) FOREST Program, No. JPMJFR206M, JSPS Grant-in-Aid for Scientific Research (B), JP 22H03663 and JSPS Grant-in-Aid for Challenging Research (Exploratory), JP 21K18781.
    I.N. thanks Research Institute for Electronic Science and Department of Mathematics at Hokkaido University for the hospitality during his sabbatical leave. 
    Y.S. was supported by JSPS Grant-in-Aid for Scientific Research (B), JP No. 21H01002. 
\end{acknowledgements}

\bibliography{reference}

\begin{thebibliography}{35}%
\makeatletter
\providecommand \@ifxundefined [1]{%
 \@ifx{#1\undefined}
}%
\providecommand \@ifnum [1]{%
 \ifnum #1\expandafter \@firstoftwo
 \else \expandafter \@secondoftwo
 \fi
}%
\providecommand \@ifx [1]{%
 \ifx #1\expandafter \@firstoftwo
 \else \expandafter \@secondoftwo
 \fi
}%
\providecommand \natexlab [1]{#1}%
\providecommand \enquote  [1]{``#1''}%
\providecommand \bibnamefont  [1]{#1}%
\providecommand \bibfnamefont [1]{#1}%
\providecommand \citenamefont [1]{#1}%
\providecommand \href@noop [0]{\@secondoftwo}%
\providecommand \href [0]{\begingroup \@sanitize@url \@href}%
\providecommand \@href[1]{\@@startlink{#1}\@@href}%
\providecommand \@@href[1]{\endgroup#1\@@endlink}%
\providecommand \@sanitize@url [0]{\catcode `\\12\catcode `\$12\catcode
  `\&12\catcode `\#12\catcode `\^12\catcode `\_12\catcode `\%12\relax}%
\providecommand \@@startlink[1]{}%
\providecommand \@@endlink[0]{}%
\providecommand \url  [0]{\begingroup\@sanitize@url \@url }%
\providecommand \@url [1]{\endgroup\@href {#1}{\urlprefix }}%
\providecommand \urlprefix  [0]{URL }%
\providecommand \Eprint [0]{\href }%
\providecommand \doibase [0]{https://doi.org/}%
\providecommand \selectlanguage [0]{\@gobble}%
\providecommand \bibinfo  [0]{\@secondoftwo}%
\providecommand \bibfield  [0]{\@secondoftwo}%
\providecommand \translation [1]{[#1]}%
\providecommand \BibitemOpen [0]{}%
\providecommand \bibitemStop [0]{}%
\providecommand \bibitemNoStop [0]{.\EOS\space}%
\providecommand \EOS [0]{\spacefactor3000\relax}%
\providecommand \BibitemShut  [1]{\csname bibitem#1\endcsname}%
\let\auto@bib@innerbib\@empty
\bibitem [{\citenamefont {Creech}\ \emph {et~al.}(2022)\citenamefont {Creech},
  \citenamefont {Guidi},\ and\ \citenamefont {Elburn}}]{creech2022artemis}%
  \BibitemOpen
  \bibfield  {author} {\bibinfo {author} {\bibfnamefont {S.}~\bibnamefont
  {Creech}}, \bibinfo {author} {\bibfnamefont {J.}~\bibnamefont {Guidi}},\ and\
  \bibinfo {author} {\bibfnamefont {D.}~\bibnamefont {Elburn}},\ }\bibfield
  {title} {\bibinfo {title} {{Artemis}: An overview of {NASA}'s activities to
  return humans to the moon},\ }in\ \href@noop {} {\emph {\bibinfo {booktitle}
  {2022 IEEE Aerospace Conference (AERO), Big Sky, MT, USA}}}\ (\bibinfo {year}
  {IEEE, New York, 2022})\ pp.\ \bibinfo {pages} {1--7}\BibitemShut {NoStop}%
\bibitem [{\citenamefont {Ikenaga}\ \emph {et~al.}(2020)\citenamefont
  {Ikenaga}, \citenamefont {Yamanaka}, \citenamefont {Ueda},\ and\
  \citenamefont {Ishii}}]{ikenaga2020study}%
  \BibitemOpen
  \bibfield  {author} {\bibinfo {author} {\bibfnamefont {T.}~\bibnamefont
  {Ikenaga}}, \bibinfo {author} {\bibfnamefont {K.}~\bibnamefont {Yamanaka}},
  \bibinfo {author} {\bibfnamefont {S.}~\bibnamefont {Ueda}},\ and\ \bibinfo
  {author} {\bibfnamefont {N.}~\bibnamefont {Ishii}},\ }\bibfield  {title}
  {\bibinfo {title} {Study on the low-energy ballistic lunar transfer orbit for
  future cargo mission to gateway},\ }in\ \href
  {https://doi.org/10.2514/6.2020-0794} {\emph {\bibinfo {booktitle} {AIAA
  Scitech 2020 Forum}}}\ (\bibinfo {year} {AIAA, Reston, VA, 2020})\BibitemShut
  {NoStop}%
\bibitem [{\citenamefont {Bandyopadhyay}\ \emph {et~al.}(2016)\citenamefont
  {Bandyopadhyay}, \citenamefont {Foust}, \citenamefont {Subramanian},
  \citenamefont {Chung},\ and\ \citenamefont
  {Hadaegh}}]{Bandyopadhyay2016review}%
  \BibitemOpen
  \bibfield  {author} {\bibinfo {author} {\bibfnamefont {S.}~\bibnamefont
  {Bandyopadhyay}}, \bibinfo {author} {\bibfnamefont {R.}~\bibnamefont
  {Foust}}, \bibinfo {author} {\bibfnamefont {G.~P.}\ \bibnamefont
  {Subramanian}}, \bibinfo {author} {\bibfnamefont {S.-J.}\ \bibnamefont
  {Chung}},\ and\ \bibinfo {author} {\bibfnamefont {F.~Y.}\ \bibnamefont
  {Hadaegh}},\ }\bibfield  {title} {\bibinfo {title} {Review of formation
  flying and constellation missions using nanosatellites},\ }\href
  {https://doi.org/10.2514/1.A33291} {\bibfield  {journal} {\bibinfo  {journal}
  {J. Spacecr. Rockets}\ }\textbf {\bibinfo {volume} {53}},\ \bibinfo {pages}
  {567} (\bibinfo {year} {2016})}\BibitemShut {NoStop}%
\bibitem [{\citenamefont {Rom-Kedar}\ and\ \citenamefont
  {Wiggins}(1990)}]{rom1990transport}%
  \BibitemOpen
  \bibfield  {author} {\bibinfo {author} {\bibfnamefont {V.}~\bibnamefont
  {Rom-Kedar}}\ and\ \bibinfo {author} {\bibfnamefont {S.}~\bibnamefont
  {Wiggins}},\ }\bibfield  {title} {\bibinfo {title} {Transport in
  two-dimensional maps},\ }\href@noop {} {\bibfield  {journal} {\bibinfo
  {journal} {Arch. Ration. Mech. Anal.}\ }\textbf {\bibinfo {volume} {109}},\
  \bibinfo {pages} {239} (\bibinfo {year} {1990})}\BibitemShut {NoStop}%
\bibitem [{\citenamefont {Prussing}\ and\ \citenamefont
  {Conway}(1993)}]{prussing1993orbital}%
  \BibitemOpen
  \bibfield  {author} {\bibinfo {author} {\bibfnamefont {J.~E.}\ \bibnamefont
  {Prussing}}\ and\ \bibinfo {author} {\bibfnamefont {B.~A.}\ \bibnamefont
  {Conway}},\ }\href@noop {} {\emph {\bibinfo {title} {Orbital mechanics}}}\
  (\bibinfo  {publisher} {Oxford University Press, Oxford, U.K.},\ \bibinfo
  {year} {1993})\BibitemShut {NoStop}%
\bibitem [{\citenamefont {Koon}\ \emph {et~al.}(2000)\citenamefont {Koon},
  \citenamefont {Lo}, \citenamefont {Marsden},\ and\ \citenamefont
  {Ross}}]{koon2000heteroclinic}%
  \BibitemOpen
  \bibfield  {author} {\bibinfo {author} {\bibfnamefont {W.~S.}\ \bibnamefont
  {Koon}}, \bibinfo {author} {\bibfnamefont {M.~W.}\ \bibnamefont {Lo}},
  \bibinfo {author} {\bibfnamefont {J.~E.}\ \bibnamefont {Marsden}},\ and\
  \bibinfo {author} {\bibfnamefont {S.~D.}\ \bibnamefont {Ross}},\ }\bibfield
  {title} {\bibinfo {title} {Heteroclinic connections between periodic orbits
  and resonance transitions in celestial mechanics},\ }\href
  {https://doi.org/10.1063/1.166509} {\bibfield  {journal} {\bibinfo  {journal}
  {Chaos}\ }\textbf {\bibinfo {volume} {10}},\ \bibinfo {pages} {427} (\bibinfo
  {year} {2000})}\BibitemShut {NoStop}%
\bibitem [{\citenamefont {Koon}\ \emph {et~al.}(2011)\citenamefont {Koon},
  \citenamefont {Lo},\ and\ \citenamefont {Ross}}]{rossDynamicalSystems}%
  \BibitemOpen
  \bibfield  {author} {\bibinfo {author} {\bibfnamefont {W.~S.}\ \bibnamefont
  {Koon}}, \bibinfo {author} {\bibfnamefont {M.~W.}\ \bibnamefont {Lo}},\ and\
  \bibinfo {author} {\bibfnamefont {S.~D.}\ \bibnamefont {Ross}},\ }\href@noop
  {} {\emph {\bibinfo {title} {Dynamical Systems, the Three-Body Problem and
  Space Mission Design}}}\ (\bibinfo  {publisher} {Marsden Books},\ \bibinfo
  {year} {2011})\BibitemShut {NoStop}%
\bibitem [{\citenamefont {Parker}\ and\ \citenamefont
  {Anderson}(2014)}]{parker2014}%
  \BibitemOpen
  \bibfield  {author} {\bibinfo {author} {\bibfnamefont {J.~S.}\ \bibnamefont
  {Parker}}\ and\ \bibinfo {author} {\bibfnamefont {R.~L.}\ \bibnamefont
  {Anderson}},\ }\href@noop {} {\emph {\bibinfo {title} {Low-Energy Lunar
  Trajectory Design}}}\ (\bibinfo  {publisher} {Wiley, Hoboken, NJ},\ \bibinfo
  {year} {2014})\BibitemShut {NoStop}%
\bibitem [{\citenamefont {McCarthy}\ and\ \citenamefont
  {Howell}(2023)}]{MCCARTHY2023556}%
  \BibitemOpen
  \bibfield  {author} {\bibinfo {author} {\bibfnamefont {B.~P.}\ \bibnamefont
  {McCarthy}}\ and\ \bibinfo {author} {\bibfnamefont {K.~C.}\ \bibnamefont
  {Howell}},\ }\bibfield  {title} {\bibinfo {title} {Four-body cislunar
  quasi-periodic orbits and their application to ballistic lunar transfer
  design},\ }\href@noop {} {\bibfield  {journal} {\bibinfo  {journal} {Adv.
  Space Res.}\ }\textbf {\bibinfo {volume} {71}},\ \bibinfo {pages} {556}
  (\bibinfo {year} {2023})}\BibitemShut {NoStop}%
\bibitem [{\citenamefont {Ross}\ \emph {et~al.}(2003)\citenamefont {Ross},
  \citenamefont {Koon}, \citenamefont {Lo},\ and\ \citenamefont
  {Marsden}}]{ross2003design}%
  \BibitemOpen
  \bibfield  {author} {\bibinfo {author} {\bibfnamefont {S.~D.}\ \bibnamefont
  {Ross}}, \bibinfo {author} {\bibfnamefont {W.~S.}\ \bibnamefont {Koon}},
  \bibinfo {author} {\bibfnamefont {M.~W.}\ \bibnamefont {Lo}},\ and\ \bibinfo
  {author} {\bibfnamefont {J.~E.}\ \bibnamefont {Marsden}},\ }\bibfield
  {title} {\bibinfo {title} {Design of a multi-moon orbiter},\ }\href@noop {}
  {\bibfield  {journal} {\bibinfo  {journal} {Adv. Astronaut. Sci.}\ }\textbf
  {\bibinfo {volume} {114}},\ \bibinfo {pages} {669} (\bibinfo {year}
  {2003})}\BibitemShut {NoStop}%
\bibitem [{\citenamefont {Ott}\ \emph {et~al.}(1990)\citenamefont {Ott},
  \citenamefont {Grebogi},\ and\ \citenamefont {Yorke}}]{ott1990controlling}%
  \BibitemOpen
  \bibfield  {author} {\bibinfo {author} {\bibfnamefont {E.}~\bibnamefont
  {Ott}}, \bibinfo {author} {\bibfnamefont {C.}~\bibnamefont {Grebogi}},\ and\
  \bibinfo {author} {\bibfnamefont {J.~A.}\ \bibnamefont {Yorke}},\ }\bibfield
  {title} {\bibinfo {title} {Controlling chaos},\ }\href@noop {} {\bibfield
  {journal} {\bibinfo  {journal} {Phys. Rev. Lett.}\ }\textbf {\bibinfo
  {volume} {64}},\ \bibinfo {pages} {1196} (\bibinfo {year}
  {1990})}\BibitemShut {NoStop}%
\bibitem [{\citenamefont {Pyragas}(1992)}]{pyragas1992continuous}%
  \BibitemOpen
  \bibfield  {author} {\bibinfo {author} {\bibfnamefont {K.}~\bibnamefont
  {Pyragas}},\ }\bibfield  {title} {\bibinfo {title} {Continuous control of
  chaos by self-controlling feedback},\ }\href@noop {} {\bibfield  {journal}
  {\bibinfo  {journal} {Phys. Lett. A}\ }\textbf {\bibinfo {volume} {170}},\
  \bibinfo {pages} {421} (\bibinfo {year} {1992})}\BibitemShut {NoStop}%
\bibitem [{\citenamefont {Yamaguti}(1994)}]{yamaguti1994towards}%
  \BibitemOpen
  \bibfield  {author} {\bibinfo {author} {\bibfnamefont {M.}~\bibnamefont
  {Yamaguti}},\ }\href@noop {} {\emph {\bibinfo {title} {Towards the Harnessing
  of Chaos}}}\ (\bibinfo  {publisher} {Elsevier, Amsterdam},\ \bibinfo {year}
  {1994})\BibitemShut {NoStop}%
\bibitem [{\citenamefont {Jaeger}\ and\ \citenamefont
  {Haas}(2004)}]{jaeger2004harnessing}%
  \BibitemOpen
  \bibfield  {author} {\bibinfo {author} {\bibfnamefont {H.}~\bibnamefont
  {Jaeger}}\ and\ \bibinfo {author} {\bibfnamefont {H.}~\bibnamefont {Haas}},\
  }\bibfield  {title} {\bibinfo {title} {Harnessing nonlinearity: Predicting
  chaotic systems and saving energy in wireless communication},\ }\href
  {https://doi.org/10.1126/science.1091277} {\bibfield  {journal} {\bibinfo
  {journal} {Science}\ }\textbf {\bibinfo {volume} {304}},\ \bibinfo {pages}
  {78} (\bibinfo {year} {2004})}\BibitemShut {NoStop}%
\bibitem [{\citenamefont {Shinbrot}\ \emph {et~al.}(1990)\citenamefont
  {Shinbrot}, \citenamefont {Ott}, \citenamefont {Grebogi},\ and\ \citenamefont
  {Yorke}}]{shinbrot1990using}%
  \BibitemOpen
  \bibfield  {author} {\bibinfo {author} {\bibfnamefont {T.}~\bibnamefont
  {Shinbrot}}, \bibinfo {author} {\bibfnamefont {E.}~\bibnamefont {Ott}},
  \bibinfo {author} {\bibfnamefont {C.}~\bibnamefont {Grebogi}},\ and\ \bibinfo
  {author} {\bibfnamefont {J.~A.}\ \bibnamefont {Yorke}},\ }\bibfield  {title}
  {\bibinfo {title} {Using chaos to direct trajectories to targets},\
  }\href@noop {} {\bibfield  {journal} {\bibinfo  {journal} {Phys. Rev. Lett.}\
  }\textbf {\bibinfo {volume} {65}},\ \bibinfo {pages} {3215} (\bibinfo {year}
  {1990})}\BibitemShut {NoStop}%
\bibitem [{\citenamefont {Bollt}\ and\ \citenamefont
  {Meiss}(1995)}]{bollt1995targeting}%
  \BibitemOpen
  \bibfield  {author} {\bibinfo {author} {\bibfnamefont {E.~M.}\ \bibnamefont
  {Bollt}}\ and\ \bibinfo {author} {\bibfnamefont {J.~D.}\ \bibnamefont
  {Meiss}},\ }\bibfield  {title} {\bibinfo {title} {Targeting chaotic orbits to
  the moon through recurrence},\ }\href@noop {} {\bibfield  {journal} {\bibinfo
   {journal} {Phys. Lett. A}\ }\textbf {\bibinfo {volume} {204}},\ \bibinfo
  {pages} {373} (\bibinfo {year} {1995})}\BibitemShut {NoStop}%
\bibitem [{\citenamefont {Schroer}\ and\ \citenamefont
  {Ott}(1997)}]{schroer1997targeting}%
  \BibitemOpen
  \bibfield  {author} {\bibinfo {author} {\bibfnamefont {C.~G.}\ \bibnamefont
  {Schroer}}\ and\ \bibinfo {author} {\bibfnamefont {E.}~\bibnamefont {Ott}},\
  }\bibfield  {title} {\bibinfo {title} {Targeting in hamiltonian systems that
  have mixed regular/chaotic phase spaces},\ }\href@noop {} {\bibfield
  {journal} {\bibinfo  {journal} {Chaos}\ }\textbf {\bibinfo {volume} {7}},\
  \bibinfo {pages} {512} (\bibinfo {year} {1997})}\BibitemShut {NoStop}%
\bibitem [{\citenamefont {Senatore}\ and\ \citenamefont
  {Ross}(2008)}]{senatore2008fuel}%
  \BibitemOpen
  \bibfield  {author} {\bibinfo {author} {\bibfnamefont {C.}~\bibnamefont
  {Senatore}}\ and\ \bibinfo {author} {\bibfnamefont {S.~D.}\ \bibnamefont
  {Ross}},\ }\bibfield  {title} {\bibinfo {title} {Fuel-efficient navigation in
  complex flows},\ }in\ \href@noop {} {\emph {\bibinfo {booktitle} {Proceedings
  of the 2008 American Control Conference}}}\ (\bibinfo {year} {IEEE, New York,
  2008})\ pp.\ \bibinfo {pages} {1244--1248}\BibitemShut {NoStop}%
\bibitem [{\citenamefont {Ramos}\ \emph {et~al.}(2018)\citenamefont {Ramos},
  \citenamefont {Garc{\'\i}a-Garrido}, \citenamefont {Mancho}, \citenamefont
  {Wiggins}, \citenamefont {Coca}, \citenamefont {Glenn}, \citenamefont
  {Schofield}, \citenamefont {Kohut}, \citenamefont {Aragon}, \citenamefont
  {Kerfoot} \emph {et~al.}}]{ramos2018lagrangian}%
  \BibitemOpen
  \bibfield  {author} {\bibinfo {author} {\bibfnamefont {A.~G.}\ \bibnamefont
  {Ramos}}, \bibinfo {author} {\bibfnamefont {V.~J.}\ \bibnamefont
  {Garc{\'\i}a-Garrido}}, \bibinfo {author} {\bibfnamefont {A.~M.}\
  \bibnamefont {Mancho}}, \bibinfo {author} {\bibfnamefont {S.}~\bibnamefont
  {Wiggins}}, \bibinfo {author} {\bibfnamefont {J.}~\bibnamefont {Coca}},
  \bibinfo {author} {\bibfnamefont {S.}~\bibnamefont {Glenn}}, \bibinfo
  {author} {\bibfnamefont {O.}~\bibnamefont {Schofield}}, \bibinfo {author}
  {\bibfnamefont {J.}~\bibnamefont {Kohut}}, \bibinfo {author} {\bibfnamefont
  {D.}~\bibnamefont {Aragon}}, \bibinfo {author} {\bibfnamefont
  {J.}~\bibnamefont {Kerfoot}}, \emph {et~al.},\ }\bibfield  {title} {\bibinfo
  {title} {Lagrangian coherent structure assisted path planning for
  transoceanic autonomous underwater vehicle missions},\ }\href@noop {}
  {\bibfield  {journal} {\bibinfo  {journal} {Sci. Rep.}\ }\textbf {\bibinfo
  {volume} {8}},\ \bibinfo {pages} {4575} (\bibinfo {year} {2018})}\BibitemShut
  {NoStop}%
\bibitem [{\citenamefont {Krishna}\ \emph {et~al.}(2022)\citenamefont
  {Krishna}, \citenamefont {Song},\ and\ \citenamefont
  {Brunton}}]{krishna2022finite}%
  \BibitemOpen
  \bibfield  {author} {\bibinfo {author} {\bibfnamefont {K.}~\bibnamefont
  {Krishna}}, \bibinfo {author} {\bibfnamefont {Z.}~\bibnamefont {Song}},\ and\
  \bibinfo {author} {\bibfnamefont {S.~L.}\ \bibnamefont {Brunton}},\
  }\bibfield  {title} {\bibinfo {title} {Finite-horizon, energy-efficient
  trajectories in unsteady flows},\ }\href
  {https://doi.org/10.1098/rspa.2021.0255} {\bibfield  {journal} {\bibinfo
  {journal} {Proc. R. Soc. London, Ser. A}\ }\textbf {\bibinfo {volume}
  {478}},\ \bibinfo {pages} {20210255} (\bibinfo {year} {2022})}\BibitemShut
  {NoStop}%
\bibitem [{\citenamefont {Haller}(2023)}]{haller2023transport}%
  \BibitemOpen
  \bibfield  {author} {\bibinfo {author} {\bibfnamefont {G.}~\bibnamefont
  {Haller}},\ }\href@noop {} {\emph {\bibinfo {title} {Transport Barriers and
  Coherent Structures in Flow Data}}}\ (\bibinfo  {publisher} {Cambridge
  University Press, Cambridge, U.K.},\ \bibinfo {year} {2023})\BibitemShut
  {NoStop}%
\bibitem [{\citenamefont {Duan}\ and\ \citenamefont
  {Wiggins}(1997)}]{duan1997lagrangian}%
  \BibitemOpen
  \bibfield  {author} {\bibinfo {author} {\bibfnamefont {J.}~\bibnamefont
  {Duan}}\ and\ \bibinfo {author} {\bibfnamefont {S.}~\bibnamefont {Wiggins}},\
  }\bibfield  {title} {\bibinfo {title} {Lagrangian transport and chaos in the
  near wake of the flow around an obstacle: a numerical implementation of lobe
  dynamics},\ }\href@noop {} {\bibfield  {journal} {\bibinfo  {journal}
  {Nonlinear Process. Geophys.}\ }\textbf {\bibinfo {volume} {4}},\ \bibinfo
  {pages} {125} (\bibinfo {year} {1997})}\BibitemShut {NoStop}%
\bibitem [{\citenamefont {Coulliette}\ and\ \citenamefont
  {Wiggins}(2001)}]{coulliette2001intergyre}%
  \BibitemOpen
  \bibfield  {author} {\bibinfo {author} {\bibfnamefont {C.}~\bibnamefont
  {Coulliette}}\ and\ \bibinfo {author} {\bibfnamefont {S.}~\bibnamefont
  {Wiggins}},\ }\bibfield  {title} {\bibinfo {title} {Intergyre transport in a
  wind-driven, quasigeostrophic double gyre: An application of lobe dynamics},\
  }\href {https://doi.org/10.5194/npg-8-69-2001} {\bibfield  {journal}
  {\bibinfo  {journal} {Nonlinear Process. Geophys.}\ }\textbf {\bibinfo
  {volume} {8}},\ \bibinfo {pages} {69} (\bibinfo {year} {2001})}\BibitemShut
  {NoStop}%
\bibitem [{\citenamefont {Lekien}\ \emph {et~al.}(2007)\citenamefont {Lekien},
  \citenamefont {Shadden},\ and\ \citenamefont
  {Marsden}}]{Lekien2007lagrangian}%
  \BibitemOpen
  \bibfield  {author} {\bibinfo {author} {\bibfnamefont {F.}~\bibnamefont
  {Lekien}}, \bibinfo {author} {\bibfnamefont {S.~C.}\ \bibnamefont
  {Shadden}},\ and\ \bibinfo {author} {\bibfnamefont {J.~E.}\ \bibnamefont
  {Marsden}},\ }\bibfield  {title} {\bibinfo {title} {Lagrangian coherent
  structures in $n$-dimensional systems},\ }\href
  {https://doi.org/10.1063/1.2740025} {\bibfield  {journal} {\bibinfo
  {journal} {J. Math. Phys.}\ }\textbf {\bibinfo {volume} {48}},\ \bibinfo
  {pages} {065404} (\bibinfo {year} {2007})}\BibitemShut {NoStop}%
\bibitem [{\citenamefont {Oshima}\ and\ \citenamefont
  {Yanao}(2015)}]{oshima2015jumping}%
  \BibitemOpen
  \bibfield  {author} {\bibinfo {author} {\bibfnamefont {K.}~\bibnamefont
  {Oshima}}\ and\ \bibinfo {author} {\bibfnamefont {T.}~\bibnamefont {Yanao}},\
  }\bibfield  {title} {\bibinfo {title} {Jumping mechanisms of trojan asteroids
  in the planar restricted three- and four-body problems},\ }\href@noop {}
  {\bibfield  {journal} {\bibinfo  {journal} {Celest. Mech. Dyn. Astron.}\
  }\textbf {\bibinfo {volume} {122}},\ \bibinfo {pages} {53} (\bibinfo {year}
  {2015})}\BibitemShut {NoStop}%
\bibitem [{\citenamefont {Naik}\ \emph {et~al.}(2017)\citenamefont {Naik},
  \citenamefont {Lekien},\ and\ \citenamefont {Ross}}]{naik2017computational}%
  \BibitemOpen
  \bibfield  {author} {\bibinfo {author} {\bibfnamefont {S.}~\bibnamefont
  {Naik}}, \bibinfo {author} {\bibfnamefont {F.}~\bibnamefont {Lekien}},\ and\
  \bibinfo {author} {\bibfnamefont {S.~D.}\ \bibnamefont {Ross}},\ }\bibfield
  {title} {\bibinfo {title} {Computational method for phase space transport
  with applications to lobe dynamics and rate of escape},\ }\href@noop {}
  {\bibfield  {journal} {\bibinfo  {journal} {Regular Chaotic Dyn.}\ }\textbf
  {\bibinfo {volume} {22}},\ \bibinfo {pages} {272} (\bibinfo {year}
  {2017})}\BibitemShut {NoStop}%
\bibitem [{\citenamefont {Chirikov}(1979)}]{CHIRIKOV1979263}%
  \BibitemOpen
  \bibfield  {author} {\bibinfo {author} {\bibfnamefont {B.~V.}\ \bibnamefont
  {Chirikov}},\ }\bibfield  {title} {\bibinfo {title} {A universal instability
  of many-dimensional oscillator systems},\ }\href@noop {} {\bibfield
  {journal} {\bibinfo  {journal} {Phys. Rep.}\ }\textbf {\bibinfo {volume}
  {52}},\ \bibinfo {pages} {263} (\bibinfo {year} {1979})}\BibitemShut
  {NoStop}%
\bibitem [{\citenamefont {Szebehely}(1967)}]{szebehely1967theories}%
  \BibitemOpen
  \bibfield  {author} {\bibinfo {author} {\bibfnamefont {V.}~\bibnamefont
  {Szebehely}},\ }\href@noop {} {\emph {\bibinfo {title} {Theories of Orbits:
  The Restricted Problem of Three Bodies}}}\ (\bibinfo  {publisher} {Academic
  Press, New York},\ \bibinfo {year} {1967})\BibitemShut {NoStop}%
\bibitem [{\citenamefont {Rom-Kedar}\ \emph {et~al.}(1990)\citenamefont
  {Rom-Kedar}, \citenamefont {Leonard},\ and\ \citenamefont
  {Wiggins}}]{rom1990analytical}%
  \BibitemOpen
  \bibfield  {author} {\bibinfo {author} {\bibfnamefont {V.}~\bibnamefont
  {Rom-Kedar}}, \bibinfo {author} {\bibfnamefont {A.}~\bibnamefont {Leonard}},\
  and\ \bibinfo {author} {\bibfnamefont {S.}~\bibnamefont {Wiggins}},\
  }\bibfield  {title} {\bibinfo {title} {An analytical study of transport,
  mixing and chaos in an unsteady vortical flow},\ }\href
  {https://doi.org/10.1017/S0022112090000167} {\bibfield  {journal} {\bibinfo
  {journal} {J. Fluid Mech.}\ }\textbf {\bibinfo {volume} {214}},\ \bibinfo
  {pages} {347} (\bibinfo {year} {1990})}\BibitemShut {NoStop}%
\bibitem [{\citenamefont {MacKay}\ \emph {et~al.}(1987)\citenamefont {MacKay},
  \citenamefont {Meiss},\ and\ \citenamefont {Percival}}]{MACKAY19871}%
  \BibitemOpen
  \bibfield  {author} {\bibinfo {author} {\bibfnamefont {R.~S.}\ \bibnamefont
  {MacKay}}, \bibinfo {author} {\bibfnamefont {J.~D.}\ \bibnamefont {Meiss}},\
  and\ \bibinfo {author} {\bibfnamefont {I.~C.}\ \bibnamefont {Percival}},\
  }\bibfield  {title} {\bibinfo {title} {Resonances in area-preserving maps},\
  }\href {https://doi.org/https://doi.org/10.1016/0167-2789(87)90002-9}
  {\bibfield  {journal} {\bibinfo  {journal} {Physica D}\ }\textbf {\bibinfo
  {volume} {27}},\ \bibinfo {pages} {1} (\bibinfo {year} {1987})}\BibitemShut
  {NoStop}%
\bibitem [{\citenamefont {Meiss}(2015)}]{meiss2015thirty}%
  \BibitemOpen
  \bibfield  {author} {\bibinfo {author} {\bibfnamefont {J.~D.}\ \bibnamefont
  {Meiss}},\ }\bibfield  {title} {\bibinfo {title} {Thirty years of turnstiles
  and transport},\ }\href {https://doi.org/10.1063/1.4915831} {\bibfield
  {journal} {\bibinfo  {journal} {Chaos}\ }\textbf {\bibinfo {volume} {25}},\
  \bibinfo {pages} {097602} (\bibinfo {year} {2015})}\BibitemShut {NoStop}%
\bibitem [{\citenamefont {MacKay}\ \emph {et~al.}(1984)\citenamefont {MacKay},
  \citenamefont {Meiss},\ and\ \citenamefont {Percival}}]{MACKAY1984transport}%
  \BibitemOpen
  \bibfield  {author} {\bibinfo {author} {\bibfnamefont {R.~S.}\ \bibnamefont
  {MacKay}}, \bibinfo {author} {\bibfnamefont {J.~D.}\ \bibnamefont {Meiss}},\
  and\ \bibinfo {author} {\bibfnamefont {I.~C.}\ \bibnamefont {Percival}},\
  }\bibfield  {title} {\bibinfo {title} {Transport in hamiltonian systems},\
  }\href {https://doi.org/https://doi.org/10.1016/0167-2789(84)90270-7}
  {\bibfield  {journal} {\bibinfo  {journal} {Physica D}\ }\textbf {\bibinfo
  {volume} {13}},\ \bibinfo {pages} {55} (\bibinfo {year} {1984})}\BibitemShut
  {NoStop}%
\bibitem [{SM()}]{SM}%
  \BibitemOpen
  \href@noop {} {}\bibinfo {note} {See Supplemental Material at [URL] for
  detailed explanations of the optimization in the standard map and Hill's
  equation and for the coordinate transformation for the periapsis Poincar{\'e}
  map in Delaunay elements}\BibitemShut {NoStop}%
\bibitem [{\citenamefont {Wiggins}(1990)}]{WIGGINS1990471}%
  \BibitemOpen
  \bibfield  {author} {\bibinfo {author} {\bibfnamefont {S.}~\bibnamefont
  {Wiggins}},\ }\bibfield  {title} {\bibinfo {title} {On the geometry of
  transport in phase space i. transport in $k$-degree-of-freedom hamiltonian
  systems, $2 \leq k < \infty$},\ }\href
  {https://doi.org/https://doi.org/10.1016/0167-2789(90)90159-M} {\bibfield
  {journal} {\bibinfo  {journal} {Physica D}\ }\textbf {\bibinfo {volume}
  {44}},\ \bibinfo {pages} {471} (\bibinfo {year} {1990})}\BibitemShut
  {NoStop}%
\bibitem [{\citenamefont {Dellnitz}\ \emph {et~al.}(2005)\citenamefont
  {Dellnitz}, \citenamefont {Junge}, \citenamefont {Koon}, \citenamefont
  {Lekien}, \citenamefont {Lo}, \citenamefont {Marsden}, \citenamefont
  {Padberg}, \citenamefont {Preis}, \citenamefont {Ross},\ and\ \citenamefont
  {Thiere}}]{dellnitz2005transport}%
  \BibitemOpen
  \bibfield  {author} {\bibinfo {author} {\bibfnamefont {M.}~\bibnamefont
  {Dellnitz}}, \bibinfo {author} {\bibfnamefont {O.}~\bibnamefont {Junge}},
  \bibinfo {author} {\bibfnamefont {W.~S.}\ \bibnamefont {Koon}}, \bibinfo
  {author} {\bibfnamefont {F.}~\bibnamefont {Lekien}}, \bibinfo {author}
  {\bibfnamefont {M.~W.}\ \bibnamefont {Lo}}, \bibinfo {author} {\bibfnamefont
  {J.~E.}\ \bibnamefont {Marsden}}, \bibinfo {author} {\bibfnamefont
  {K.}~\bibnamefont {Padberg}}, \bibinfo {author} {\bibfnamefont
  {R.}~\bibnamefont {Preis}}, \bibinfo {author} {\bibfnamefont {S.~D.}\
  \bibnamefont {Ross}},\ and\ \bibinfo {author} {\bibfnamefont
  {B.}~\bibnamefont {Thiere}},\ }\bibfield  {title} {\bibinfo {title}
  {Transport in dynamical astronomy and multibody problems},\ }\href@noop {}
  {\bibfield  {journal} {\bibinfo  {journal} {Int. J. Bifurcat. Chaos}\
  }\textbf {\bibinfo {volume} {15}},\ \bibinfo {pages} {699} (\bibinfo {year}
  {2005})}\BibitemShut {NoStop}%
\end{thebibliography}%
\end{document}


\title{Supplemental Material for ``Designing robust trajectories by lobe dynamics\\ in low-dimensional Hamiltonian systems''}

\author{Naoki Hiraiwa}
\affiliation{Department of Aeronautics and Astronautics, Kyushu University, 744 Motooka, Nishi-ku, Fukuoka 8190395, Japan}
\affiliation{Colorado Center for Astrodynamics Research, University of Colorado Boulder, 3775 Discovery
Drive, Boulder, Colorado 80303, USA}
\author{Mai Bando}
\affiliation{Department of Aeronautics and Astronautics, Kyushu University, 744 Motooka, Nishi-ku, Fukuoka 8190395, Japan}
\author{Isaia Nisoli}
\affiliation{Instituto de Matem{\'a}tica, Universidade Federal do Rio de Janeiro, Avenida Athos da Silveira Ramos, 149, Edif{\'i}cio do Centro de Tecnologia, Bloco C (T{\'e}rreo), Cidade Universit{\'a}ria 21941-909, Brazil}
\author{Yuzuru Sato}
\email{ysato@math.sci.hokudai.ac.jp}
\affiliation{RIES-MSC/Department of Mathematics, Hokkaido University, N12 W7 Kita-ku, Sapporo, 0600812 Hokkaido, Japan}
\affiliation{London Mathematical Laboratory, 14 Buckingham Street, London WC2N 6DF, United Kingdom}
\date{\today}

\maketitle

The supplemental material contains:
\begin{itemize}
   \item Supplementary figures for effective lobe sequences in the standard map
   \item Illustration of the optimization in the standard map and Hill's equation
   \item Coordinate transformation for the periapsis Poincar{\'e} map in Delaunay elements
\end{itemize}

\section{Effective lobe sequences in the standard map}
\begin{figure}[b]
    \centering
    \includegraphics[width=0.95\columnwidth]{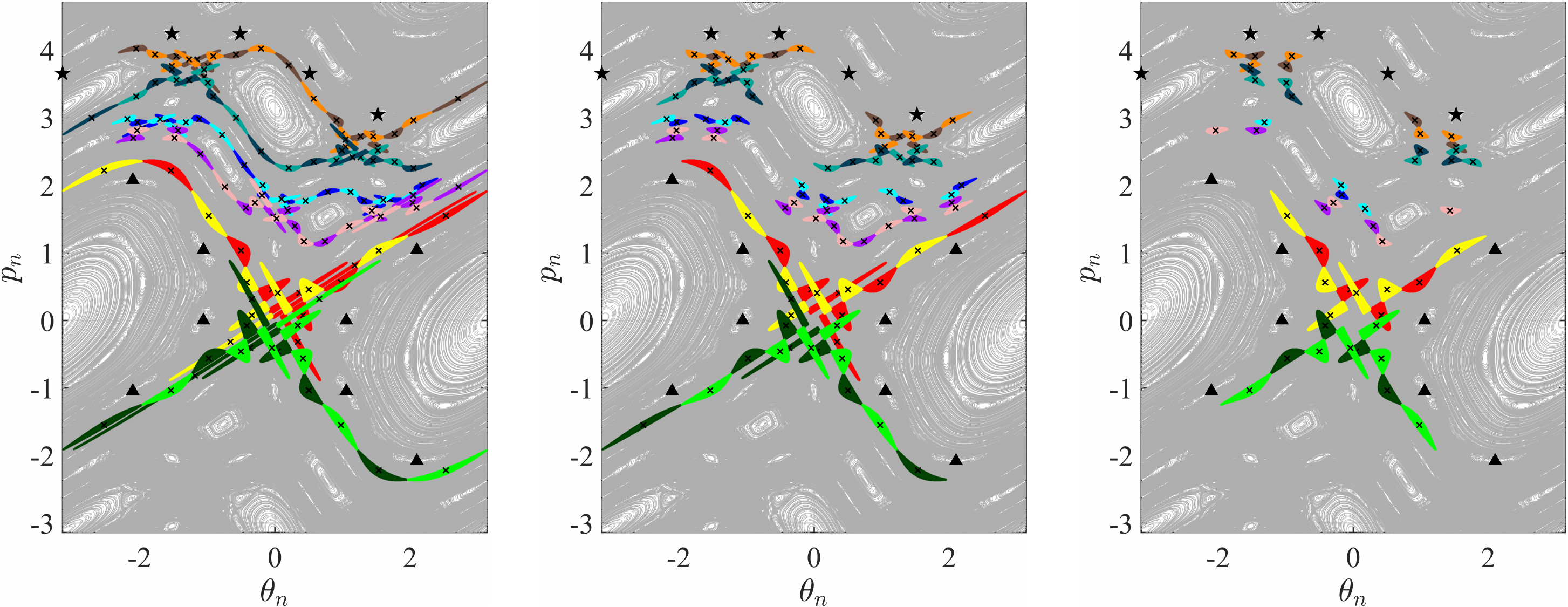}
    \caption{\label{fig:effective_lobe_supp} Selected $12$ effective lobe sequences $S(L_i,r^*)\ (i = 1,2,\ldots,12)$ when $r^*=0.01$ (left), $r^*=0.02$ (middle), and $r^*=0.03$ (right) in the standard map of $K=1.2$. Different $S(L_i,r^*)$ is colored differently. The potential start points, potential goal points, and centers of gravity of the effective lobes are denoted as triangles, stars, and crosses, respectively.}
\end{figure}
In the following, $S(L,r^*)$ denotes an effective lobe sequence starting from a lobe $L$ and composed of lobes whose radius is larger than $r^*$.
Figure~\ref{fig:effective_lobe_supp} illustrates the selected $12$ effective lobe sequences $S(L_i,r^*)\ (i = 1,2,\ldots,12)$, each with distinct values of $r^*$ in the standard map of $K=1.2$.
Near the unstable fixed point at $(p_n, \theta_n) = (0, 0)$, the effective lobes in dark green, light green, red, and yellow represent $S(L_1,r^*)$, $S(L_2,r^*)$, $S(L_3,r^*)$, and $S(L_4,r^*)$, respectively.
Similarly, in the vicinity of the unstable periodic orbit with period $3$, the effective lobes in purple, light pink, blue, and cyan correspond to $S(L_5,r^*)$, $S(L_6,r^*)$, $S(L_7,r^*)$, and $S(L_8,r^*)$, respectively.
Finally, around the unstable periodic orbit with period $2$, the effective lobes in blue--green, royal blue, brown, and orange are associated with $S(L_9,r^*)$, $S(L_{10},r^*)$, $S(L_{11},r^*)$, and $S(L_{12},r^*)$, respectively.
Generally, a larger $r^*$ results in the inclusion of fewer lobes within each effective lobe sequence.
While other lobe sequences can be identified by focusing on different pairs of the stable/unstable manifolds of the saddles, these $12$ lobe sequences are selected owing to their relatively larger lobe radii than those of the other lobes around the same periodic orbits.

\section{Optimization in the standard map}
\begin{figure}[b]
    \begin{minipage}{\columnwidth}
        \centering
        \includegraphics[width=0.9\columnwidth]{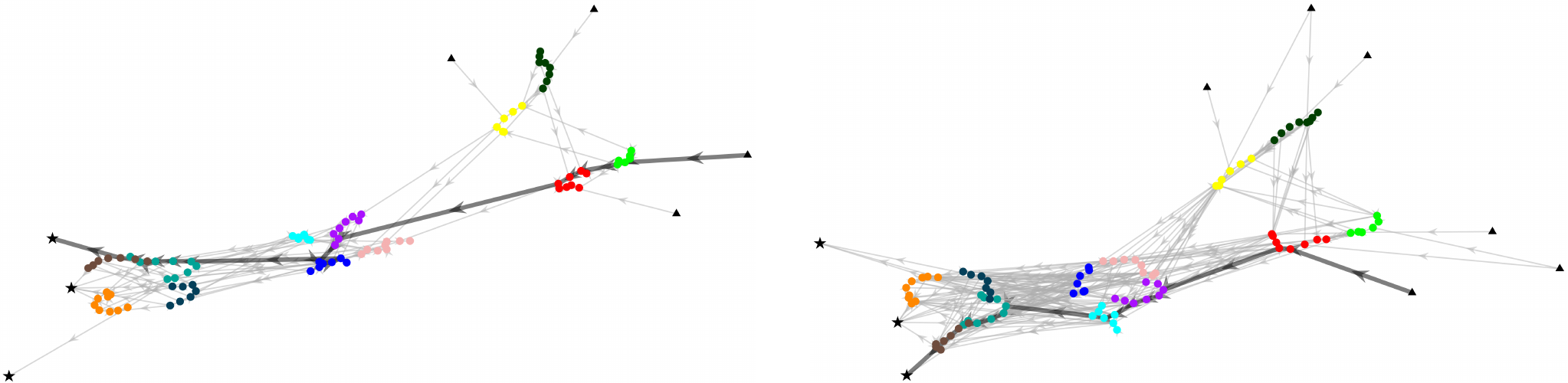}
        \caption{\label{fig:optimal_path_standard_map} Optimal paths for $r^*=0.02$, and $d^* = 0.64$ (left) and $d^*=1.07$ (right) in the standard map of $K=1.2$. The edge length is proportional to the edge weight. The potential start points, potential goal points, and centers of gravity of the effective lobes are denoted as triangles, stars, and colored dots, respectively. These colored dots correspond to the colored regions in Fig.~\ref{fig:effective_lobe_supp}. A bold line is the optimal path.}
    \end{minipage}\\%
    \vspace{3mm}
    \begin{minipage}{\columnwidth}
        \centering
        \includegraphics[width=0.9\columnwidth]{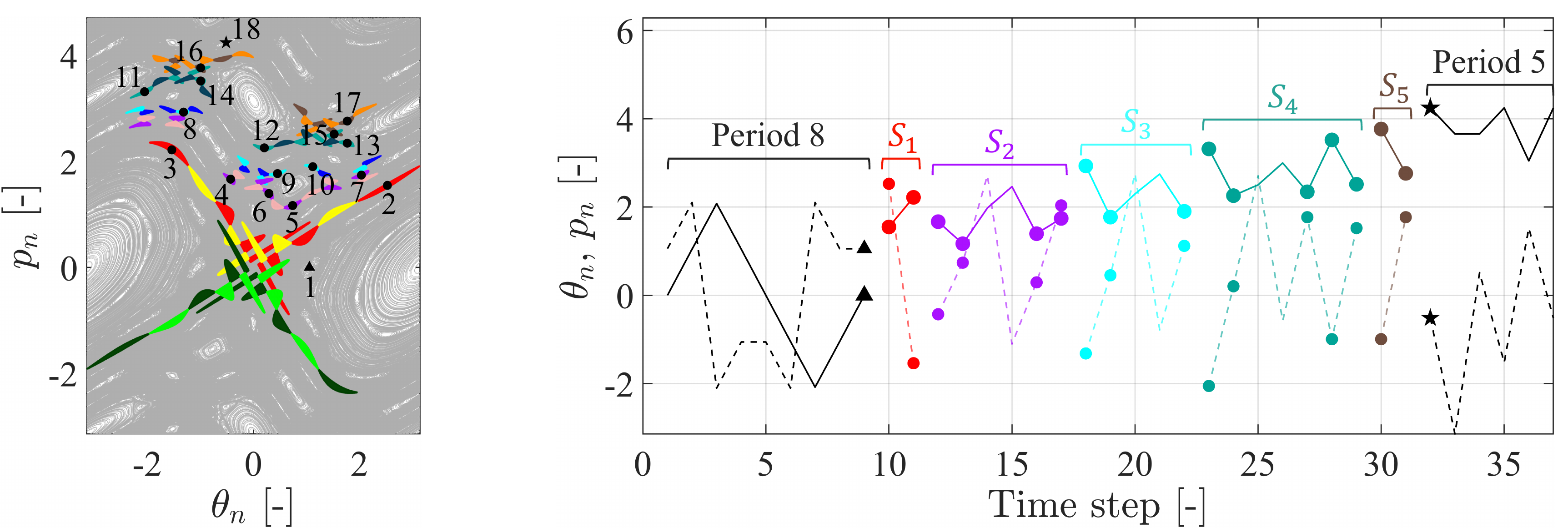}
        \caption{\label{fig:time_history_supp} Optimal trajectory for the standard map with $K = 1.2$, with a total cost $D = 1.9633$ and transfer time $\Delta n = 23$, where $r^*=0.02$ and $d^*=1.07$: State space of the standard map (left), and controlled time series of $p_n$ [solid line] and $\theta_n$ [dashed line] (right) are depicted. Gaps on the right panel indicate controlled jumps. Different effective lobe sequences are colored differently. The start point, goal point, and centers of gravity of the adopted lobes are denoted as a triangle, a star, and dots, respectively. The numbers in the left panel represent the order of transfer, similar to those in Fig.~$2$.}
    \end{minipage}
\end{figure}%
In the present control scheme, we search for the optimal trajectory based on the graph that reflects all the selected potential transfer paths under $0 < d_n = |\Delta p_n| < d^*$.
In the example of Fig.~$3$, we assume that transfer is possible from the $8$ start points to $S(L_i,r^*)\ (i = 1,2,3,4)$, from $S(L_i,r^*)\ (i = 1,2,3,4)$ to $S(L_j,r^*)\ (j = 5,6,7,8)$, from $S(L_j,r^*)\ (j = 5,6,7,8)$ to $S(L_k,r^*)\ (k = 9,10,11,12)$, and from $S(L_k,r^*)\ (k = 9,10,11,12)$ to the $5$ goal points.
In addition, around each of the unstable orbits selected for finding lobe sequences, transfers from $S(L_{4m-3},r^*)$ or $S(L_{4m-2},r^*)$ to $S(L_{4m-1},r^*)$ or $S(L_{4m},r^*)$ $(m = 1,2,3)$ are also considered.
These transfers are the ones between lobe sequences around the same orbit, which is essential to rapidly overcome the partial barriers by resonances~\cite{MACKAY19871,meiss2015thirty}.
It should be noted that although connecting all the centers of gravity of the effective lobes around the same orbit is possible, it expands the search space, consequently increasing the computational cost to find the optimal trajectory.
The graphs for the optimization with $d^* = 0.64, 1.07$ are depicted in Fig.~\ref{fig:optimal_path_standard_map}.
Here, the edge length is proportional to the edge weight, with the weight of the edges between lobes within the same lobe sequence set at $0.07$ (rather than $0$) for visualization purposes.
The restriction of remaining in an effective lobe sequence for at least one unit of time is not incorporated into the graph structure, as this constraint depends on the specific paths from the start orbit to the goal orbit.
The optimal paths for $d^* = 0.64, 1.07$, identified by analyzing the graph structure, are presented in Fig.~\ref{fig:optimal_path_standard_map}. The left and right panels correspond to the optimal trajectories depicted in Figs.~$3$ and~\ref{fig:time_history_supp}, respectively. In Fig.~$3$, $S_i\ (i=1,2,\ldots,6)$ refers to $S(L_j,r^*)\ (j = 2,3,5,7,9,11)$ respectively, and in Fig.~\ref{fig:time_history_supp}, $S_k\ (k=1,2,\ldots,5)$ indicates $S(L_\ell,r^*)\ (\ell = 3,5,8,9,11)$ respectively.
The total cost is $D = 2.1333$ for $d^* = 0.64$ and $D = 1.9633$ for $d^* = 1.07$, and the transfer time is $\Delta n = 23$ for $d^* = 0.64, 1.07$. The corresponding control histories are explained in Table~\ref{tab:control_history}, where hyphens represent natural transitions on the start/goal orbits or jumps between the centers of gravity of the lobes within the same effective lobe sequence, with the associated costs presumed to be zero.
In these examples, a larger maximum jump cost $d^*$ reduces the total cost owing to the expanded search space. As illustrated in the right panel of Table~\ref{tab:control_history}, the control cost at time step $n = 11$ exceeds $0.64$, indicating that this path is not viable in the scenario presented in the left panel of the table.
\begin{table}[t]
    \centering
    \caption{Control histories when $r^*=0.02$ and $d^* = 0.64$ (left) and $d^* = 1.07$ (right) in the standard map of $K=1.2$. Hyphens indicate the state transition whose cost is presumed to be zero.}
    \label{tab:control_history}
    \begin{minipage}[t]{0.48\textwidth}
    \centering
    \begin{tabular}{>{\centering}p{1in}>{\centering}p{1in}>{\centering}p{1in}l}
        \hline\hline
        Time step & Control input & Control timing & \\
        $n$       & $\Delta p_n$  & $\eta_n$        & \\ \hline
        $1$       & -             & -              & \\
        $2$       & -             & -              & \\
        $3$       & -             & -              & \\
        $4$       & -             & -              & \\
        $5$       & -             & -              & \\
        $6$       & -             & -              & \\
        $7$       & -             & -              & \\
        $8$       & -             & -              & \\
        $9$       & $-0.50461$    & $0.84193$      & \\
        $10$      & -             & -              & \\
        $11$      & $0.15354$     & $0.50310$      & \\
        $12$      & -             & -              & \\
        $13$      & -             & -              & \\
        $14$      & $0.57065$     & $0.85845$      & \\
        $15$      & -             & -              & \\
        $16$      & $0.14556$     & $0.074673$     & \\
        $17$      & -             & -              & \\
        $18$      & -             & -              & \\
        $19$      & -             & -              & \\
        $20$      & -             & -              & \\
        $21$      & -             & -              & \\
        $22$      & -             & -              & \\
        $23$      & $0.39992$     & $0.046098$     & \\
        $24$      & -             & -              & \\
        $25$      & -             & -              & \\
        $26$      & -             & -              & \\
        $27$      & -             & -              & \\
        $28$      & -             & -              & \\
        $29$      & $0.052697$    & $1.0561\times10^{-5}$ & \\
        $30$      & -             & -              & \\
        $31$      & $0.30634$     & $0.80552$      & \\
        $32$      & -             & -              & \\
        $33$      & -             & -              & \\
        $34$      & -             & -              & \\
        $35$      & -             & -              & \\
        $36$      & -             & -              & \\
        \hline\hline
    \end{tabular}
    \end{minipage}
    \hspace{0.02\textwidth}
    \begin{minipage}[t]{0.48\textwidth}
    \centering
    \begin{tabular}{>{\centering}p{1in}>{\centering}p{1in}>{\centering}p{1in}l}
        \hline\hline
        Time step & Control input & Control timing & \\
        $n$       & $\Delta p_n$  & $\eta_n$        & \\ \hline
        $1$       & -             & -              & \\
        $2$       & -             & -              & \\
        $3$       & -             & -              & \\
        $4$       & -             & -              & \\
        $5$       & -             & -              & \\
        $6$       & -             & -              & \\
        $7$       & -             & -              & \\
        $8$       & -             & -              & \\
        $9$       & $0.50461$     & $0.15807$      & \\
        $10$      & -             & -              & \\
        $11$      & $0.64805$     & $0.86307$      & \\
        $12$      & -             & -              & \\
        $13$      & -             & -              & \\
        $14$      & -             & -              & \\
        $15$      & -             & -              & \\
        $16$      & -             & -              & \\
        $17$      & $0.11688$     & $0.022443$     & \\
        $18$      & -             & -              & \\
        $19$      & -             & -              & \\
        $20$      & -             & -              & \\
        $21$      & -             & -              & \\
        $22$      & $0.33468$     & $0.61582$      & \\
        $23$      & -             & -              & \\
        $24$      & -             & -              & \\
        $25$      & -             & -              & \\
        $26$      & -             & -              & \\
        $27$      & -             & -              & \\
        $28$      & -             & -              & \\
        $29$      & $0.052697$    & $1.0561\times10^{-5}$ & \\
        $30$      & -             & -              & \\
        $31$      & $0.30634$     & $0.80552$      & \\
        $32$      & -             & -              & \\
        $33$      & -             & -              & \\
        $34$      & -             & -              & \\
        $35$      & -             & -              & \\
        $36$      & -             & -              & \\
        \hline\hline
    \end{tabular}
    \end{minipage}
\end{table}

\section{Optimization in Hill's Equation}
\begin{figure}[t]
    \centering
    \includegraphics[width=0.95\columnwidth]{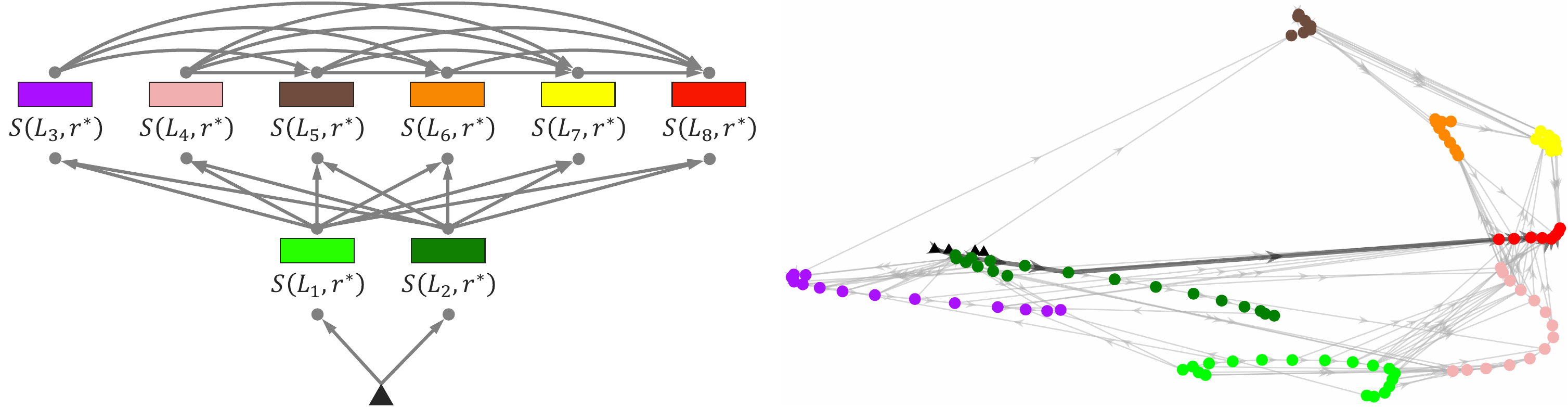}
    \caption{\label{fig:graph_CR3BP} Optimization process with $r^*=0.002$ and $d^* = 0.09760$ ($100$~[m/s]) in Hill's equation with $J = 3.16$: All potential paths for the optimization (left), and the optimal path on the graph (right) are depicted. The edge length is proportional to the edge weight. The potential start points and centers of gravity of the effective lobes are denoted as triangles and colored dots, respectively. These colored dots correspond to the colored regions in Fig.~$4$. A bold line is the optimal path.}
\end{figure}

In this section, we briefly outline the optimization process for the numerical example in Hill's equation with a Jacobi integral of $J = 3.16$, similar to the process discussed in the previous section.
Initially, all potential transfer paths are determined as depicted in the left panel of Fig.~\ref{fig:graph_CR3BP}, and the colors match those of the effective lobe sequences in the bottom panel of Fig.~$4$. We find $S(L_i,r^*)\ (i = 1,2)$ and $S(L_j,r^*)\ (j = 3,4,\ldots,8)$ around the $7$:$2$ and $3$:$1$ unstable resonant orbits, respectively. A $p$:$q$ resonant orbit is defined as an orbit whose period in the inertial frame $T$ is related to the Moon's period $T_2$ as $pT = qT_2$. In the example of Fig.~$4$, the minimum lobe radius $r^*$ is set at $r^* = 0.002$.
The maximum jump cost $d^*$ is determined to enable transfers from the periapses of the $7$:$2$ neutrally stable resonant orbit to $S(L_i,r^*)\ (i = 1,2)$ and from $S(L_i,r^*)\ (i = 1,2)$ to $S(L_j,r^*)\ (j = 3,4,\ldots,8)$, and prevent direct transfer from these periapses to $S(L_j,r^*)\ (j = 3,4,\ldots,8)$. In essence, $d^*$ must satisfy $0.07137\ldots < d^* < 0.10109\ldots$. Consequently, we set $d^* = 0.09760$ ($100$~[m/s]) in Fig.~$4$. The right panel of Fig.~\ref{fig:graph_CR3BP} displays the graph for this optimization. The edge weights between lobes within the same lobe sequence are adjusted to $1\times10^{-7}$~[m/s] instead of $0$~[m/s] for visualization purposes.
Following this methodology, we identify the optimal path for $d^* = 0.09760$ ($100$~[m/s]), as depicted in Fig.~\ref{fig:graph_CR3BP}. The corresponding optimal trajectory is represented in Fig.~$4$.

\begin{figure}[b]
    \centering
    \includegraphics[width=0.9\columnwidth]{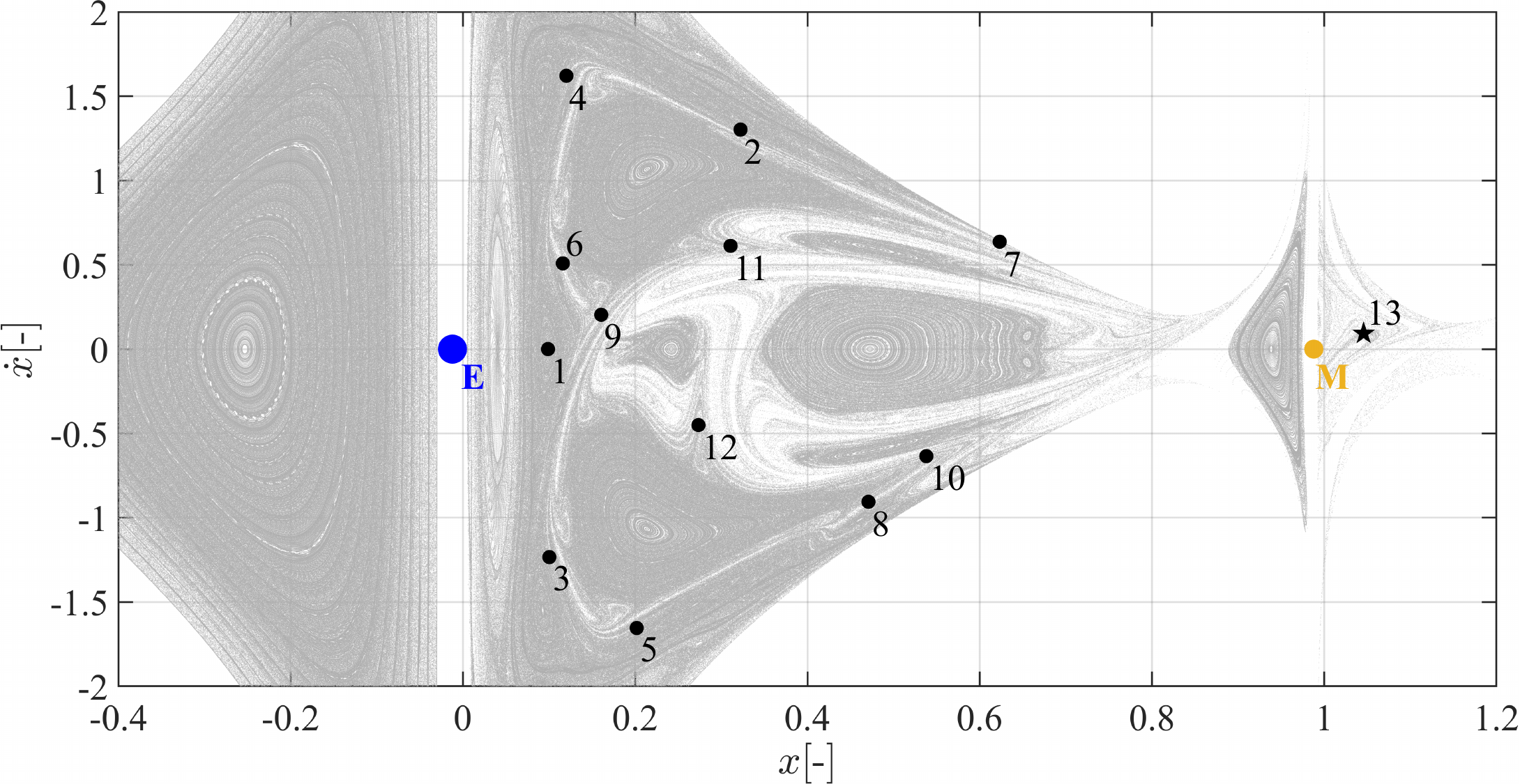}
    \caption{\label{fig:poincare_CR3BP} Optimal trajectory in Fig.~$4$ on the Poincar{\'e} section of $y = 0\ (\dot{y} > 0)$ in Hill's equation with $J = 3.16$. Blue and yellow dots represent the Earth and Moon, respectively. The optimal trajectory and goal point are denoted as black dots and a star, respectively. The numbers indicate the order of transfer.}
\end{figure}

On the other hand, prior research on spacecraft trajectory design in the Earth--Moon frame~\cite{bollt1995targeting,schroer1997targeting} has mainly focused on constructing the transfer trajectories on the Poincar{\'e} section at $y = 0\ (\dot{y} > 0)$. Figure~\ref{fig:poincare_CR3BP} presents the optimal trajectory from Fig.~$4$ on this Poincar{\'e} section. The start point of this trajectory does not appear on the surface of this section. Despite selecting a Jacobi integral value slightly different from that in the previous work~\cite{bollt1995targeting,schroer1997targeting}, Fig.~\ref{fig:poincare_CR3BP} illustrates that our control scheme effectively facilitates the placement of intermediate waypoints for transfer in a chaotic zone.

\section{Action-angle coordinates and Delaunay elements}
\begin{figure}[b]
  \centering
  \includegraphics[width=0.4\columnwidth]{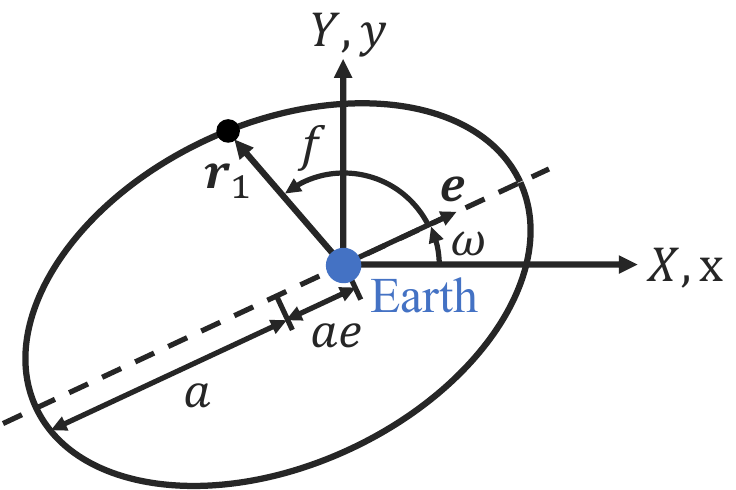}
  \caption{Classical orbital elements of an elliptic orbit in the temporary inertial frame (when $X\dot{Y} - Y\dot{X} > 0$)} \label{fig:inertial_frame}
\end{figure}

Our control scheme operates on the periapsis Poincar{\'e} map in the example with Hill's equation, as this map is a natural choice for observing lobe dynamics in this case. The map is plotted in Delaunay elements, which are canonical variables, i.e., suitable for depicting an area-preserving map. We perform the following coordinate transformation to calculate the Delaunay elements from the spacecraft state $(x,\,\,y,\,\,\dot{x},\,\,\dot{y})$. Initially, the spacecraft state in the rotating $x$--$y$ frame is converted to the temporary inertial $X$--$Y$ frame, as follows:
\begin{align}
    X &= x + \mu, & Y &= y, & \dot{X} &= \dot{x} - y, & \dot{Y} &= \dot{y} + (x+\mu)
\end{align}
On this inertial frame, the semi-major axis $a$, eccentricity $e$, argument of periapsis $\omega$, and true anomaly $f$ are defined as
\begin{align}
    a &= \frac{\left(1-\mu\right)r_1}{2\left(1-\mu\right) - r_1 V^2}, & 
    \bm{e} &= - \frac{\left(\bm{r}_1 \times \bm{V}\right) \times \bm{V}}{1-\mu} - \frac{\bm{r}_1}{r_1}, \qquad e = \|\bm{e}\|
\end{align}
$\omega$ and $f$ are the angles indicated in Fig.~\ref{fig:inertial_frame}, where $\mu$ is the parameter in Hill's equation, $\bm{r}_1 = \,^t[X,\,\,Y]$, $r_1 = \|\bm{r}_1\|$, $\bm{V} = \,^t[\dot{X},\,\,\dot{Y}]$, and $V = \|\bm{V}\|$.
The four variables $(a,\,e,\,\omega,\,f)$ are called the classical orbital elements. This transformation assumes that the angular momentum $X\dot{Y} - Y\dot{X}$ is positive. Finally, we transform the classical orbital elements $(a,\,e,\,\omega,\,f)$ to the Delaunay elements $(l_d,\,g_d,\,L_d,\,G_d)$ as follows:
\begin{align}
    l_d &= M = E - e\sin E, &
    g_d &= \omega, &
    L_d &= \sqrt{\left(1 - \mu\right)a}, &
    G_d &= \sqrt{\left(1 - \mu\right)a\left(1 - e^2\right)}
\end{align}
where $M$ denotes the mean anomaly and $E$ indicates the eccentric anomaly expressed as 
\begin{equation}
    E = 2\tan^{-1}\left(\sqrt{\frac{1 - e}{1 + e}}\tan\frac{f}{2}\right) \label{eq:trans_f2e}
\end{equation}%
The periapsis passage is expressed as $f = 0$. Based on Eq.~\eqref{eq:trans_f2e}, this condition is identical with $l_d = E - e\sin E =0$.
